\documentclass[aps,prd,preprint,superscriptaddress,amsmath,amssymb,showpacs]{revtex4-1}
\usepackage{dcolumn}
\usepackage{graphicx}
\usepackage{float}
\usepackage{physics}
\usepackage[colorlinks=true,allcolors=blue]{hyperref}
\begin{document}

\title{Chiral phase transition and spin alignment of vector mesons with chiral imbalance
	in a rotating QCD medium } 

\author{Yang~Hua}
\affiliation{%
College of Science, China Three Gorges University, Yichang 443002, China\\
}%

\author{Sheng-Qin~Feng}
\email{ Corresponding author: fengsq@ctgu.edu.cn}
\affiliation{%
College of Science, China Three Gorges University, Yichang 443002, China\\
}%
\affiliation{%
Center for Astronomy and Space Sciences and Institute of Modern Physics, China
Three Gorges University, Yichang 443002, China\\
}%
\affiliation{%
Key Laboratory of Quark and Lepton Physics (MOE) and Institute of Particle
Physics, Central China Normal University, Wuhan 430079, China\\
}%

\date{\today}%

\begin{abstract}
We study the two-flavor Nambu-Jona-Lasinio model under the rotation and chiral chemical potential $\mu_{5}$. First, the influence of chiral imbalance on the chiral phase transition in the $T_{pc}-\omega$ plane is investigated. Research manifests that as $\mu_{5}$ increases, the critical point of the $T_{pc}-\omega$ plane chiral phase transition will move closer to the $T$ axis. This means that the chiral chemical potential $\mu_{5}$ can significantly affect the $T_{pc}-\omega$ phase diagram and phase transition behavior. While discussing the $T_{pc}-\omega$ phase diagram, we also study the spin alignment of the $\rho$ vector meson under rotation. In the study of the spin alignment of the vector meson $\rho$, $\rho_{00}$ is the $00$ element of the spin density matrix of vector mesons. At high temperatures, $\rho_{00}$ is close to $1/3$, which indicates that the spin alignment of the vector meson $\rho$ is isotropic. The study found that under finite rotation, increasing the chiral chemical potential $\mu_{5}$  can significantly enhance $\rho_{00}$ around the phase transition temperature. When rotational angular velocity is zero, $\rho_{00}$ is close to $1/3$, but as $\omega$ increases, $\rho_{00}$ significantly decreases and deviates $1/3$, indicating that rotation can significantly cause polarization characteristics. The $\rho_{00}-r$  relationship near the phase transition temperature is studied. It is found that the farther away from the center of rotation, the lower the degree of spin polarization of the system. It is also found that the influence of chiral imbalance on the $\rho_{00}-r$  relationship is also significant.

\end{abstract}


\maketitle

\section{Introduction}\label{sec:Intro}
Rotation exists in various physical systems in nature. For example, astrophysical objects like neutron stars, which are composed of dense QCD matter, can rotate rapidly~\cite{Watts:2016uzu,Grenier:2015pya}. The QCD matter produced in noncentral heavy-ion collisions has a finite angular momentum of $10^{4}$-$10^{5}\hbar$, with a local angular velocity in the range of 0.01-0.1 GeV~\cite{Wang:2024szr,Wang:2024rim,Huang:2011ru,Becattini:2007sr,Csernai:2013bqa,Deng:2016gyh,Jiang:2016woz},which affects the spin and orbital angular momentum of quarks and gluons~\cite{Becattini:2015ska,Shi:2017wpk,Pang:2016igs,STAR:2017ckg,Becattini:2016gvu,Xia:2018tes}. This provides a unique opportunity to study the vacuum structure of quantum chromodynamics at hadron scales in a rotating background and the strong interaction matter. The spin-orbit coupling in QCD may lead to many interesting phenomena. Liang and Wang predicted the global spin polarization of final state hyperons~\cite{Liang:2004ph} and vector meson spin alignment~\cite{Liang:2004xn} in 2005, which was later confirmed by heavy-ion collision experiments~\cite{STAR:2017ckg,STAR:2022fan}.

The QCD vacuum possesses certain nontrivial gluon configurations, such as instantons~\cite{tHooft:1976rip} or sphalerons~\cite{tHooft:1976rip,Manton:1983nd,Klinkhamer:1984di}, which result in a nontrivial topological structure characterized by an integer topological winding number $Q_{W}$. The quark-gluon plasma (QGP) that arises in heavy-ion collisions~\cite{Kharzeev:2004ey} is a phase of extremely hot QCD matter. In this state, a significant number of topological gluon configurations, known as sphalerons, are expected to be produced. Therefore, the quark-gluon plasma is the best place to find direct experimental evidence for the existence of gauge field configurations with nonzero winding number.

In the QGP, these gluon configurations with nonzero winding numbers can be generated anywhere at any time. Once these gluon configurations are excited at a certain spacetime point, the topological charge of the vacuum around that point will be altered by these gluon configurations. The chirality imbalance is induced by the nonzero topological charge through the axial anomaly of QCD. According to the Adler-Bell-Jackiw anomaly relation, the change in chirality $N_{R} - N_{L}$ is connected to the topological winding number of the gluon field as
\begin{equation}\label{eq:1}
	\begin{aligned}
		N_{5}=N_{R}-N_{L}=-2Q_{W},
	\end{aligned}
\end{equation}
where $N_{5}$ is the chiral charge, quantifying the difference between the number of right-handed  and left-handed  quarks, while the integer $Q_{W}$ is the topological winding number of the gluon field~\cite{Fukushima:2008xe,Fukushima:2010fe,Yu:2014sla,Hou:2020zhb,Chao:2013qpa}. According to the sign of their winding numbers, they can convert left-handed quarks into right-handed quarks or vice versa through axial anomalies. In other words, each gluon topological transition with a nonzero $Q_{W}$ will induce a corresponding fluctuation in the quark chiral imbalance $N_{5}$ .
This process will result in chiral imbalance between right and left quarks, leading to the violation of parity ($P$) and charge-parity ($CP$) symmetry in the thermal plasma~\cite{Yu:2014sla,Hou:2020zhb,Chao:2013qpa}.

At the same time, spin, as an important quantum degree of freedom of quark matter, can point either upward or downward in any specific direction specified by an external probe. Chirality, vorticity, and magnetic fields achieve spin polarization and spin alignment of quark matter through their influence on quark spins~\cite{Guo:2019joy,Lv:2024uev,Chen:2015hfc}.
Therefore, it is significant to study the impact of chiral imbalance on quark spin alignment in rotating systems. To simulate the topological induced change of chiral charge $N_{5}=N_{R}-N_{L}$, we can introduce a chiral chemical potential $\mu_{5}$ to study the difference in the number of left and right quarks. Therefore, the chiral chemical potential can reflect the degree of chiral imbalance, and the greater the chiral chemical potential, the higher the corresponding degree of chiral imbalance.

The Nambu-Jona-Lasinio (NJL) model is an effective field model that has certain advantages in describing the characteristics of chiral phase transition and deconfinement phase transition in noncentral nucleus-nucleus collisions with rotation and strong magnetic fields ~\cite{Jiang:2016wvv,Ebihara:2016fwa,Chernodub:2016kxh,Liu:2017spl,Chernodub:2017ref,Wang:2018sur,Cao:2019ctl,Chen:2019tcp,Sun:2021hxo,Zhu:2022irg,Xu:2022hql,Cao:2023yba,Cao:2023olg,Qiu:2023kwv,Qiu:2023ezo,Bao:2024glw}.
The role of gluons is integrated out, and quarks do not transmit through gluons in the NJL model. By introducing local four-fermion interactions, the formal chiral symmetry of the NJL model can be preserved. Therefore, it is necessary to consider the contribution of gluon configurations, namely the chiral imbalance problem in the QGP medium, which is caused by the nonzero gluon configurations in the QCD background. This chirality imbalance becomes particularly important, especially in the case of rotation.

In high-energy heavy-ion collisions, due to the enormous orbital angular momentum, the QGP produced has a very strong vortical field, and through spin-orbit coupling, it leads to the polarization of partons. The root of this spin polarization phenomenon is the spin-orbit coupling in strong interactions. In rotating QGP, $\rho$ mesons may exhibit a spin alignment phenomenon, where their spin state is correlated with the direction of rotation of the system. This spin alignment phenomenon is related to the chiral chemical potential and angular velocity. Some studies~\cite{Liang:2004xn,STAR:2022fan,Chen:2023hnb,Sheng:2022wsy,Gao:2020lxh} have manifested that whether the spin density matrix element $\rho_{00}$ of the vector meson deviates from $1/3$ is a key factor in studying the spin alignment of the vector meson $\rho$.

In this work, we will first utilize the two-flavor NJL model to study the chiral phase diagram of  $T_{pc}-\omega$. Then we will systematically investigate the spin alignment of the vector meson $\rho$ by considering the chiral imbalance. This paper is organized as follows. The two-flavor NJL model with chiral imbalance under rotation is introduced in Sec.~\ref{sec:2}. The quark polarization as well as the spin alignment of the vector meson is introduced in Sec.~\ref{sec:3}. The detailed influences of chiral imbalance on chiral phase transition and spin alignment of the vector meson are studied in Sec.~\ref{sec:4}. The summaries and conclusions are made in Sec.~\ref{sec:5}.

\section{Nambu-Jona-Lasinio model with chiral imbalance under rotation }\label{sec:2}

We will study the chiral phase transition and spin alignment with a chiral imbalance created by sphaleron transitions under rotation in the NJL model, which is described by the Lagrangian density. The Lagrangian of the two-flavor NJL model with chiral imbalance under rotation can be written as
\begin{equation}\label{eq:(2)}
	\begin{aligned}
		\mathcal{L}_{NJL}=\bar{\psi}[i\bar{\gamma}^{\mu}(\partial_{\mu}+\Gamma_{\mu})-m+\mu\gamma^{0}+\mu_{5}\gamma^{0}\gamma^{5}]\psi+G_{s}[(\bar{\psi}\psi)^{2}+(\bar{\psi}i\gamma^{5}\vec{\tau}\psi)^{2}],
	\end{aligned}
\end{equation}
where $\psi$ represents the two-flavor quark field, $\bar{\gamma}^{\mu}=e_{a}^{\mu}\gamma^{a}$ with $e_{a}^{\mu} $ is the tetrads for spinors and $\gamma^{a} $ represents the gamma matrix. $\Gamma_{\mu}$ is defined as $\Gamma_{\mu}=\dfrac{1}{4}\times\dfrac{1}{2}[\gamma^{a},\gamma^{b}]\Gamma_{ab\mu}$ which is the spinor connection, where $\Gamma_{ab\mu}=\eta_{ac}(e_{\sigma}^{c}G_{\mu\nu}^{\sigma}e_{b}^{\nu}-e_{b}^{\nu}\partial_{\mu}e_{\nu}^{c})$ is the affine connection determined by $g^{\mu\nu}$, $e_{\mu}^{a}=\delta_{\mu}^{a}+\delta_{i}^{a}\delta_{\mu}^{0}\upsilon_{i}$ and $e_{a}^{\mu}=\delta_{a}^{\mu}-\delta_{a}^{0}\delta_{i}^{\mu}\upsilon_{i}$~\cite{Jiang:2016wvv}. There are other parameters in Eq.(2), such as the following: $m$ is the bare quark mass matrix, $\mu_{5}$ denotes the chiral chemical potential, $\mu$ denotes the chemical potential, where $G_{s}$ represents the coupling constant for the scalar-type four-quark interaction, and $\vec{\tau}$ is the Pauli matrix.  When considering a system with an angular velocity along the fixed $z$ axis, one can obtain the rotational velocity  $\vec{\upsilon}=\vec{\omega}\times\vec{x}$. Under the mean field approximation (MFA), the final Lagrangian~\cite{Jiang:2016wvv,Wang:2018sur,Xu:2022hql} is given as

\begin{equation}\label{eq:(3)}
	\begin{aligned}
		\mathcal{L}_{MFA}=&\bar{\psi}[i\gamma^{\mu}\partial_{\mu}-M+\mu\gamma^{0}+\mu_{5}\gamma^{0}\gamma^{5}+(\gamma^{0})^{-1}((\vec{\omega}\times\vec{x})\cdot(-i\vec{\partial})
\\& + \vec{\omega}\cdot\vec{S}_{4\times4})]\psi-G_{s}\sigma^{2},
	\end{aligned}
\end{equation}
where $\omega$ is the angular velocity,
$\vec{S}_{4\times4} =\dfrac{1}{2} \left(
\begin{array}{cccc}
	\vec{\sigma} & 0\\
	0 & \vec{\sigma}\\
\end{array} \right)$ is the spin operator, $\sigma=\langle\bar{\psi}\psi\rangle$ is the so-called chiral condensate and $M=m-2G_{s}\sigma$ is the effective quark mass.

First, we will briefly introduce some basic knowledge about rotating matter. For a rotating frame, the spacetime metric $g_{\mu\nu}$ is
\begin{equation}\label{eq:4}
	\begin{aligned}
		g_{\mu\nu}=\left(
		\begin{array}{cccc}
			1-\vec{\upsilon}^{2}&-\upsilon_{1}&-\upsilon_{2}&-\upsilon_{3}\\
			-\upsilon_{1}&-1&0&0\\
			-\upsilon_{2}&0&-1&0\\
			-\upsilon_{3}&0&0&-1\\
		\end{array} \right) ,\\
	\end{aligned}
\end{equation}
where $\upsilon_{i}$ represents the three components of velocity and the magnitude of velocity is $\upsilon=\sqrt{\upsilon_{1}^{2}+\upsilon_{2}^{2}+\upsilon_{3}^{2}}$.
By using the cylindrical coordinates, we can derive the general positive-energy solution  ~\cite{Xu:2022hql} for the quark field from the Dirac equation corresponding to the Lagrangian given in Eq.(3) as
\begin{equation}\label{eq:5}
	\begin{aligned}
		\mathbf{\psi}(\theta,r)=e^{-iEt+ip_{z}z}\left(
		\begin{array}{cccc}
			ce^{in\theta}J_{n}(p_{t}r)\\
			ide^{i(n+1)\theta}J_{n+1}(p_{t}r)\\
			c^{\prime}e^{in\theta}J_{n}(p_{t}r)\\
			id^{\prime}e^{i(n+1)\theta}J_{n+1}(p_{t}r)\\
		\end{array} \right) ,\\
	\end{aligned}
\end{equation}
where $J_{n}(p_{t}r)$ is the Bessel function of the first kind. The coefficients $c$, $d$, $c^{\prime}$, and $d^{\prime}$ satisfy $c^{2}+d^{2}+c^{\prime2}+d^{\prime2}=1$ required by normalization, and they also satisfy $\dfrac{d}{c}=\dfrac{d^{\prime}}{c^{\prime}}=\lambda$, where the $\lambda$ is given as
\begin{equation}\label{eq:6}
	\begin{aligned}
		\lambda=-2\mu_{5}p_{t}\big/ \big[(\sqrt{p_{t}^{2}+p_{z}^{2}}-s\mu_{5})^{2}-p_{t}^{2}-(p_{z}+\mu_{5})^{2}\big].
	\end{aligned}
\end{equation}

The dispersion relation is given as
\begin{equation}\label{eq:7}
	\begin{aligned}
		E_{n,s}=\sqrt{(\sqrt{p_{t}^{2}+p_{z}^{2}}-s\mu_{5})^{2}+M^{2}}-(n+\dfrac{1}{2})\omega,
	\end{aligned}
\end{equation}
where $s=\pm1$ and $n$ is the quantum number of angular momentum. The negative-energy solution can be obtained in a similar way. Now the grand potential density under MFA can be obtained as
\begin{equation}\label{eq:(8)}
	\begin{aligned}
		\Omega= & \dfrac{(M-m)^{2}}{4G_{s}}-\dfrac{N_{f}N_{c}}{8\pi^{2}}\sum_{n=-\infty}^\infty \sum_{s=\pm1}\int\mathrm{d}p_{t}^{2}\mathrm{d}p_{z}W_{n,s}
\\& \times\left\lbrace E_{n,s}+T\ln[1+e^{-\beta(E_{n,s}-\mu)}]+T\ln[1+e^{-\beta(E_{n,s}+\mu)}]\right\rbrace,
	\end{aligned}
\end{equation}
where $\beta=\dfrac{1}{T}$, $W_{n,s}=[J_{n}^{2}(p_{t}r)+\lambda^{2}J_{n+1}^{2}(p_{t}r)]/(1+\lambda^{2})$,  $N_{f}=2$, and $N_{c}=3$ for the two-flavor NJL model. Then, the constituent mass or chiral condensate will be required to minimize the grand thermodynamical potential. The values are determined by solving the stationary condition as
\begin{equation}\label{eq:(9)}
	\begin{aligned}
		\dfrac{\partial\Omega}{\partial M}=0,\hskip 0.6cm \dfrac{\partial^{2}\Omega}{\partial M^{2}}>0.
	\end{aligned}
\end{equation}

It can be found that the energy integral term in the thermodynamic potential Eq.(8) is ultraviolet divergent. Since the NJL model is nonrenormalizable, a cutoff needs to be introduced in the momentum integration. In this study, we use the smoothing regularization scheme by introducing a shape factor $f_{\Lambda}(p)$ in the divergent zero-point energy and choose the soft-cutoff weight function as
\begin{equation}\label{eq:(10)}
	\begin{aligned}
		f_{\Lambda}(p)=\sqrt{\dfrac{\Lambda^{2N}}{\Lambda^{2N}+ |\vec{p}\;|^{2N}}},
	\end{aligned}
\end{equation}
where $N=5$ is chosen.

\section{Spin Alignment of Vector Meson $\rho$ }\label{sec:3}

The $\rho$ meson is a type of meson belonging to the vector mesons in QCD. The vector meson $\rho$ is composed of a quark and an antiquark, with their spins parallel, corresponding to $s = 1$. In certain extreme conditions, such as in the rotating QGP produced in noncentral heavy-ion collisions, $\rho$ mesons can be generated through the production mechanism of quark-antiquark pairs. $\rho$ mesons may exhibit spin alignment phenomena, where their spin states are correlated with the rotation direction of the system.

We will briefly discuss the phenomenon of spin alignment of vector mesons under rotation. From the perspective of the simple quark model, the spin of the $\rho$ vector meson depends on the spin of its constituent quarks. The spin state of a vector meson is described by $3\times3$ the spin density matrix $\rho_{ij}$. However, due to parity conservation in the strong decay process, the polarization of vector mesons cannot be directly measured. Therefore, determining whether the spin of a meson is parallel or antiparallel to its angular momentum is a challenging task, and only through experiments $\rho_{00}$ can be obtained.

The spin alignment of vector mesons can be measured by the angular distribution of decay products in the vector meson decaying to two spinless particles ~\cite{Liang:2004xn,STAR:2022fan,Schilling:1969um}:
\begin{equation}\label{eq:11}
	\begin{aligned}
		\dfrac{\mathrm{d}N}{\mathrm{d}\cos\theta^{*}}=\dfrac{3}{4}[1-\rho_{00}+(3\rho_{00}-1)\cos^{2}\theta^{*}],
	\end{aligned}
\end{equation}
where $\theta^{*}$ is the polar angle between the quantization axis and the momentum direction of the decay particle. Thus, we need to find appropriate quantities to describe the spin density. By
considering the chiral imbalance between the spin up and spin down of quarks in the rotation background, one defines~\cite{Xu:2024kdh,Sun:2024anu} the spin polarization of the quark as
\begin{equation}\label{eq:(12)}
	\begin{aligned}
		P_{q}=\dfrac{N_{\uparrow}^{+}-N_{\downarrow}^{+}}{N_{\uparrow}^{+}+N_{\downarrow}^{+}},
	\end{aligned}
\end{equation}
where $N_{\uparrow/\downarrow}^{+/-}$ denote the quark (antiquark) number density with spin up (down), respectively, which can be extracted by taking the partial derivative of $\Omega$ Eq.(8) with respect to $\mu$. The spin-orbit coupling method is obtained through energy eigenstates that possess a certain total angular momentum. The detailed expressions for $N_{\uparrow}^{+}$, $N_{\downarrow}^{+}$, $N_{\uparrow}^{-}$ and $N_{\downarrow}^{-}$ are listed as
\begin{equation}\label{eq:(13)}
	\begin{aligned}
		N_{\uparrow}^{+}=\dfrac{N_{f}N_{c}}{4\pi^{2}}\sum_{n=-\infty}^\infty \sum_{s=\pm1}\int\mathrm{d}p_{t}\mathrm{d}p_{z} p_{t}\dfrac{J_{n}^{2}(p_{t}r)}{1+\lambda^{2}}\dfrac{e^{-\beta(E_{n,s}-\mu)}}{1+e^{-\beta(E_{n,s}-\mu)}},
	\end{aligned}
\end{equation}
\begin{equation}\label{eq:(14)}
	\begin{aligned}
		N_{\downarrow}^{+}=\dfrac{N_{f}N_{c}}{4\pi^{2}}\sum_{n=-\infty}^\infty \sum_{s=\pm1}\int\mathrm{d}p_{t}\mathrm{d}p_{z} p_{t}\dfrac{\lambda^{2}J_{n+1}^{2}(p_{t}r)}{1+\lambda^{2}}\dfrac{e^{-\beta(E_{n,s}-\mu)}}{1+e^{-\beta(E_{n,s}-\mu)}},
	\end{aligned}
\end{equation}
\begin{equation}\label{eq:(15)}
	\begin{aligned}
		N_{\uparrow}^{-}=-\dfrac{N_{f}N_{c}}{4\pi^{2}}\sum_{n=-\infty}^\infty \sum_{s=\pm1}\int\mathrm{d}p_{t}\mathrm{d}p_{z} p_{t}\dfrac{J_{n}^{2}(p_{t}r)}{1+\lambda^{2}}\dfrac{e^{-\beta(E_{n,s}+\mu)}}{1+e^{-\beta(E_{n,s}+\mu)}},
	\end{aligned}
\end{equation}
\begin{equation}\label{eq:(16)}
	\begin{aligned}
		N_{\downarrow}^{-}=-\dfrac{N_{f}N_{c}}{4\pi^{2}}\sum_{n=-\infty}^\infty \sum_{s=\pm1}\int\mathrm{d}p_{t}\mathrm{d}p_{z} p_{t}\dfrac{\lambda^{2}J_{n+1}^{2}(p_{t}r)}{1+\lambda^{2}}\dfrac{e^{-\beta(E_{n,s}+\mu)}}{1+e^{-\beta(E_{n,s}+\mu)}}.
	\end{aligned}
\end{equation}

By using a simple recombination model~\cite{Liang:2004xn} of polarized quarks and antiquarks, one can postulate that the $\rho$ vector meson is generated by the simple recombination of quarks and antiquarks with polarization $P_{q}$ and $P_{\bar{q}}$, respectively. The spin alignment of vector meson is taken as
\begin{equation}\label{eq:(17)}
	\begin{aligned}
		\rho_{00}=\dfrac{1-P_{q}P_{\bar{q}}}{3+P_{q}P_{\bar{q}}};
	\end{aligned}
\end{equation}
when $P_{q}P_{\bar{q}}$ is small, the spin alignment of $\rho$ can be approximated as
\begin{equation}\label{eq:(18)}
	\begin{aligned}
		\rho_{00}\approx\dfrac{1}{3}-\dfrac{4}{9}P_{q}P_{\bar{q}};
	\end{aligned}
\end{equation}
then, the spin alignment of $\rho$ can be approximately expressed as
\begin{equation}\label{eq:19}
	\begin{aligned}
		\rho_{00}=\dfrac{1}{3}-\dfrac{4}{9}\dfrac{N_{\uparrow}^{+}-N_{\downarrow}^{+}}{N_{\uparrow}^{+}
+N_{\downarrow}^{+}}\dfrac{N_{\uparrow}^{-}-N_{\downarrow}^{-}}{N_{\uparrow}^{-}+N_{\downarrow}^{-}}.
	\end{aligned}
\end{equation}

This article adopts the recombination model proposed by Liang and Wang~\cite{Liang:2004xn}, suggesting that vector mesons $\rho$ can be produced through the recombination of polarized quarks and antiquarks, where the spin polarizations of the quarks and antiquarks are $P_{q}$ and $P_{\bar{q}}$, respectively.

In studying the spin alignment of vector mesons, besides the recombination model, there is also the quark condensation model~\cite{Yang:2017sdk}. The spin alignment of vector mesons takes the following form:
\begin{equation}\label{eq:(20)}
	\begin{aligned}
		\rho_{00}= \frac{1}{3}-\frac{4}{9}(\beta \omega)^{2},
	\end{aligned}
\end{equation}
where $\beta=1/T$.

According to the NJL model study in~\cite{Wei:2023pdf} with $T$ = 150 MeV, the spin alignment of vector mesons (including $\rho$ and $\phi$) manifests the following relationship:
\begin{equation}\label{eq:(21)}
	\begin{aligned}
		\rho_{00}= \frac{1}{3}-5.10\omega^{2}+39.62\omega^{4},
	\end{aligned}
\end{equation}
This model can be regarded as a self-consistent model for studying spin alignment under rotation using the NJL model. It is worth mentioning that there are also theories ~\cite{cao:2020pmm,sheng:2022ssp} that use the self-consistent  model to study spin alignment under strong magnetic fields using the NJL model.
\begin{figure}[ht]
	\centering
	\includegraphics[width=.60\textwidth]{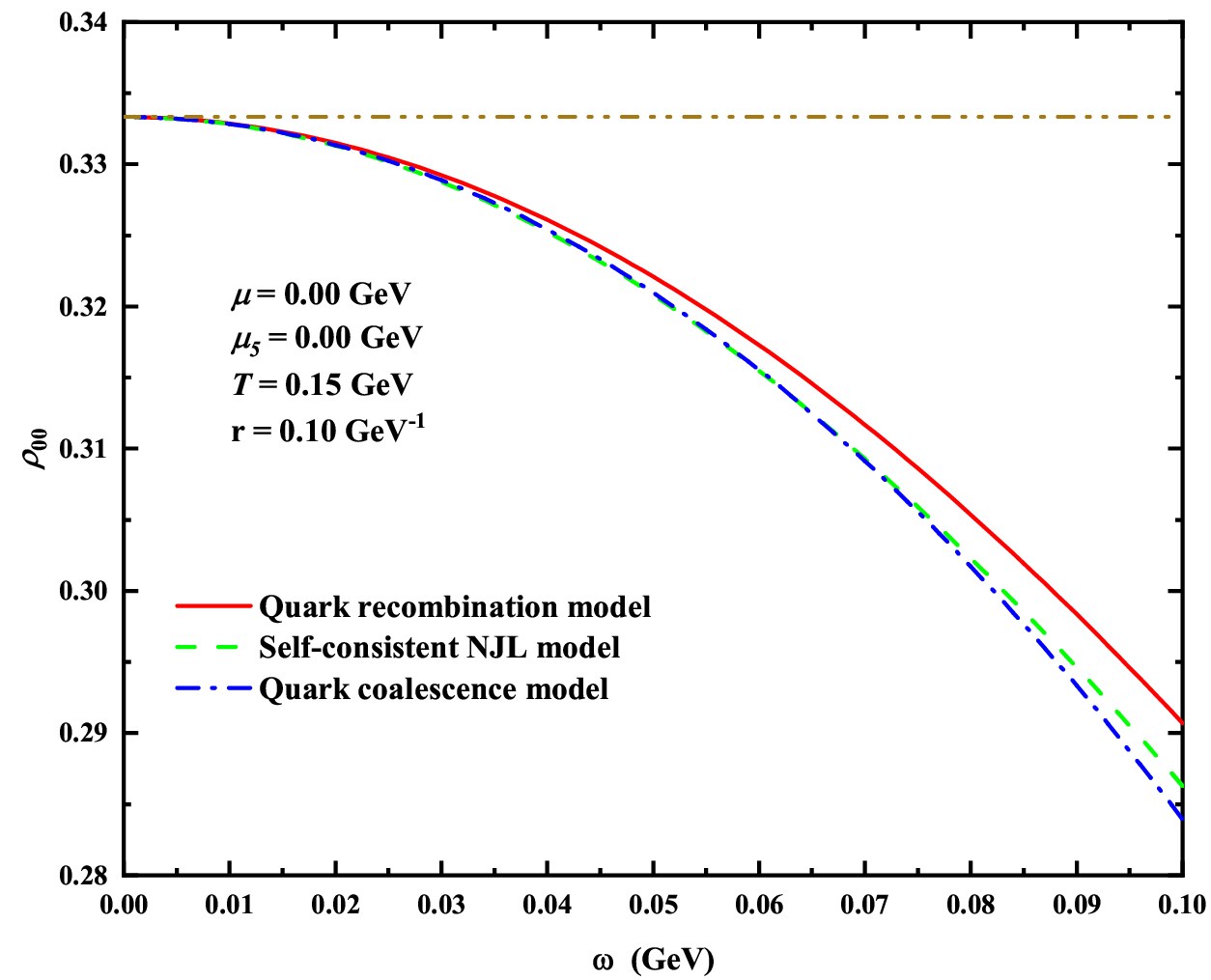}
	\caption{The $\rho$ meson spin alignment as a function of angular velocity $\omega$ at $T$ = 0.15 GeV, $\mu$ = 0.0 GeV, and $\mu_{5}$ = 0.0 GeV in the recombination
model and comparing with the quark coalescence model and self-consistent NJL model.}
	\label{fig:1}
\end{figure}

In Fig.~\ref{fig:1}, we show the variation of spin alignment of $\rho$ mesons with $\omega$ in the recombination model, self-consistent NJL model, and quark condensate model. It is noteworthy that under rotation, the spin alignment of vector mesons exhibits a negative deviation of $\rho_{00}-1/3$ . As can be seen from Fig.~\ref{fig:1}, the deviation of spin alignment of vector mesons in the recombination model from $1/3$ is less significant.

\section{NUMERICAL RESULTS AND DISCUSSIONS}\label{sec:4}

To calibrate sets of parameters to applicable observables, parameters for the NJL model under rotation are chosen as $m=0.006$ GeV, $\Lambda=626.76$ MeV, and $G_{s}\Lambda^{2}=2.02$, as reported in Ref.~\cite{Yu:2014sla}. The empirical values are given as $f_{\pi}=92.3$ MeV, the vacuum quark condensate $\langle \bar{u}u\rangle^{1/3}=-251$ MeV and the constituent quark mass $ M = 325$ MeV. Additionally, the $z$-angular-momentum quantum number is denoted by $n=0,\pm1,\pm2,\ldots$. In principle, the sum of all values of $n$ is essential. However, the rapid convergence of these expressions allows us to restrict the summation of $n$ from -5 to 5~\cite{Jiang:2016wvv,Sun:2023kuu,Sun:2024anu}. It naturally has the limitations  $\omega r < 1$ in our calculation due to the limitation of the speed of light.
\begin{figure}[ht]
	\centering
	\includegraphics[width=.60\textwidth]{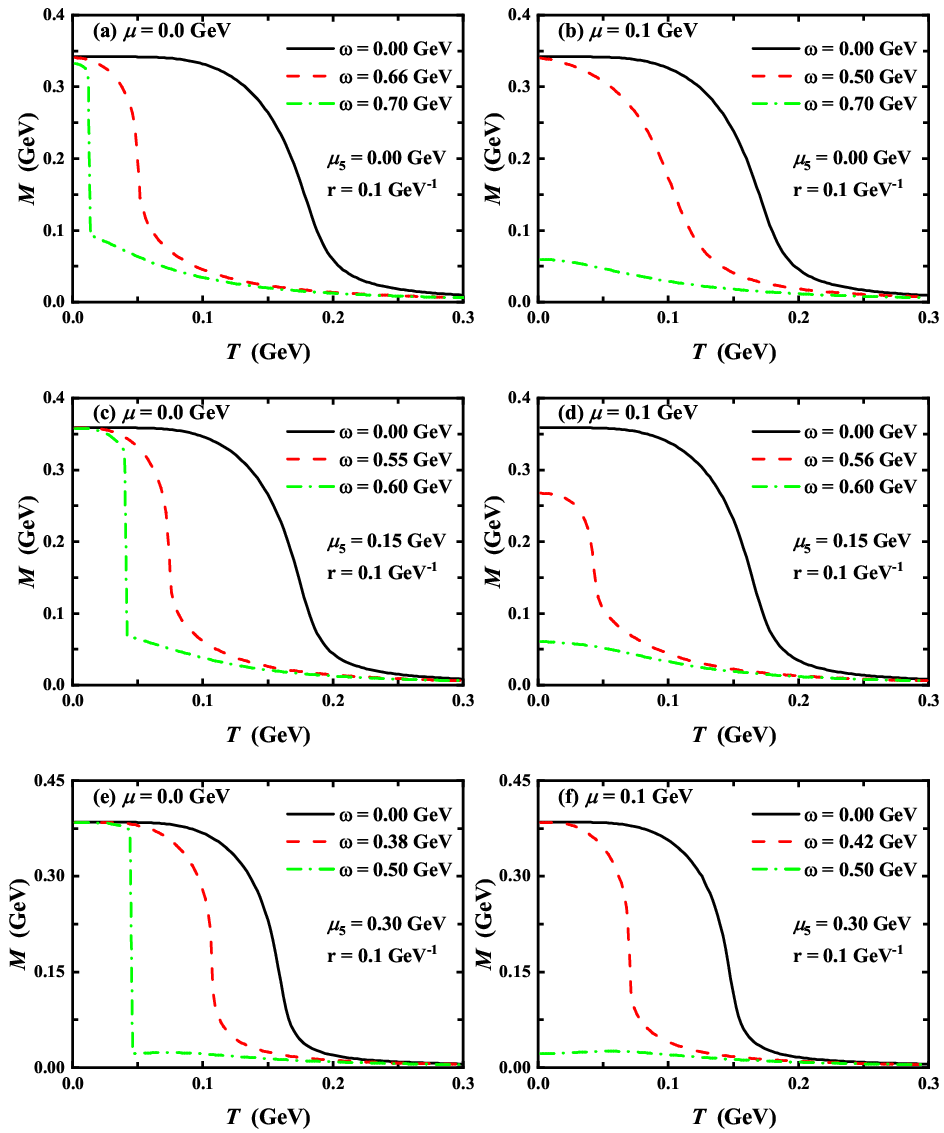}
	\caption{The dependence of dynamical quark mass ($M$) on temperature ($T$) for different angular
		velocity with different chiral chemical potential $\mu_{5}$. (a), (c) and (e) are for $\mu = 0.0$ GeV, and (b), (d) and (f)
		are for $\mu = 0.1$ GeV, respectively.}
	\label{fig:2}
\end{figure}

In these six $M-T$ diagrams, the smooth black solid line shows the $M-T$ relationship at angular velocity $\omega = 0$, corresponding to the second-order phase transition process. The red dashed line manifests the $ M-T$ characteristic of the critical end point (CEP), and the green dotted line shows the first-order phase transition with abrupt changes. Figures~\ref{fig:2}(a),~\ref{fig:2}(c), and ~\ref{fig:2}(e) show the $M-T$ diagram when the chemical potential $\mu$ is equal to $0$, while Figs.~\ref{fig:2}(b), ~\ref{fig:2}(d), and ~\ref{fig:2}(f) show the $M-T$ diagram when the chemical potential is equal to $0.1$ GeV. When $\mu = 0$, it can be seen that at lower temperatures, the value of $M$ increases with the increase of the chiral chemical potential. The critical temperature corresponding to the CEP (red dashed line) increases with the increase of the chiral chemical potential, while the critical angular velocity decreases with the increase of the chiral chemical potential.

When the chemical potential $\mu = 0.1$ GeV and the chiral chemical potential $\mu_{5} = 0$ [corresponding to Fig.~\ref{fig:2}(b)], even if the rotational angular velocity is increased, the red dashed line in Fig.~\ref{fig:2}(b) still exhibits a smooth variation, corresponding to a second-order phase transition. That is to say, when the chemical potential $\mu = 0.1$ GeV and the chiral chemical potential $\mu_{5} = 0$ GeV, the system undergoes only a second-order phase transition without CEP or a first-order phase transition. Let us focus on the green dashed lines in Figs.~\ref{fig:2}(b), ~\ref{fig:2}(d), and ~\ref{fig:2}(f) ($\mu = 0.1$ GeV), which correspond to larger chemical potentials and larger rotational angular velocities. It is found that the value of $M$ is very small and almost does not change with temperature. At this time, the chiral condensation is almost zero, indicating that the system has entered the stage of chiral symmetry restoration. By comparing the $M-T$ curves at zero chemical potential and $\mu = 0.1$ GeV, it can be seen that, as the chemical potential increases, the critical temperature of CEP decreases, but the critical angular velocity of CEP increases, which is exactly the opposite of the effect of chiral chemical potential on CEP.

\begin{figure}[ht]
	\centering
	\includegraphics[width=.35\textwidth]{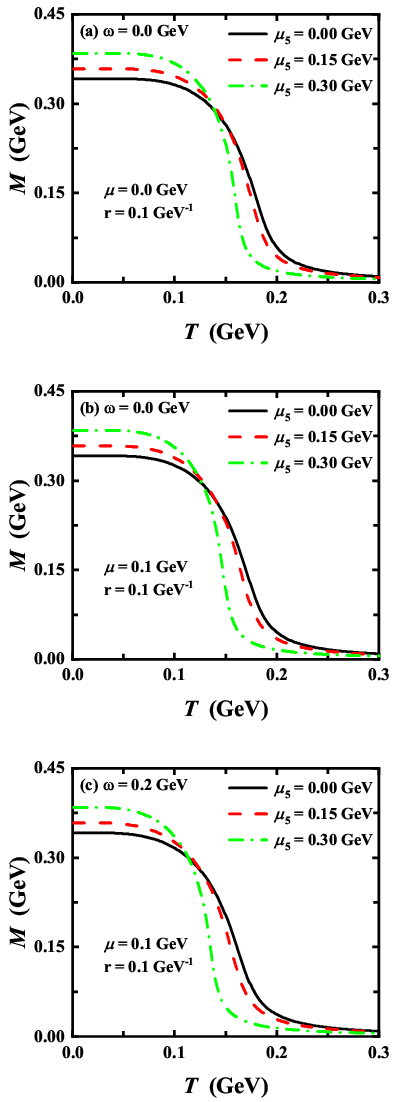}
	\caption{The dependence of dynamical quark mass ($M$) on temperature ($T$) for different chiral chemical potential $\mu_{5}$.
(a) is for angular velocity $\omega$ = 0 GeV, and chemical potential $\mu$ = 0 GeV. (b) is for angular velocity $\omega$ = 0 GeV, and chemical potential $\mu$ = 0.1 GeV.
(c) is for angular velocity $\omega$ = 0.2 GeV, and chemical potential $\mu$ = 0.1 GeV.}
	\label{fig:3}
\end{figure}

The dependence of dynamical quark mass ($M$) on temperature ($T$) for different chiral chemical potentials are manifested in Fig.~\ref{fig:3}. It is found that, even when the rotation angular velocity and chemical potential both are zero [as shown in Fig.~\ref{fig:3}(a)], $\mu_{5}$  can reinforce chiral symmetry breaking at low temperature. However, as the temperature increases, the situation becomes completely different. When the temperature rises near or above the phase transition temperature, increasing the chiral chemical potential $\mu_{5}$  will favor the restoration of chiral symmetry, which is consistent with the findings in Refs.~\cite{Chernodub:2011fr,Fukushima:2010fe}. References~\cite{Cao:2014uva,Cao:2015xja} primarily discuss chiral condensation at zero temperature, and they discovered that increasing the chiral chemical potential $\mu_{5}$ reinforces chiral symmetry breaking, which is consistent with our results.

\subsection{Phase diagram}\label{4A}

\begin{figure}[ht]
	\centering
	\includegraphics[width=\textwidth]{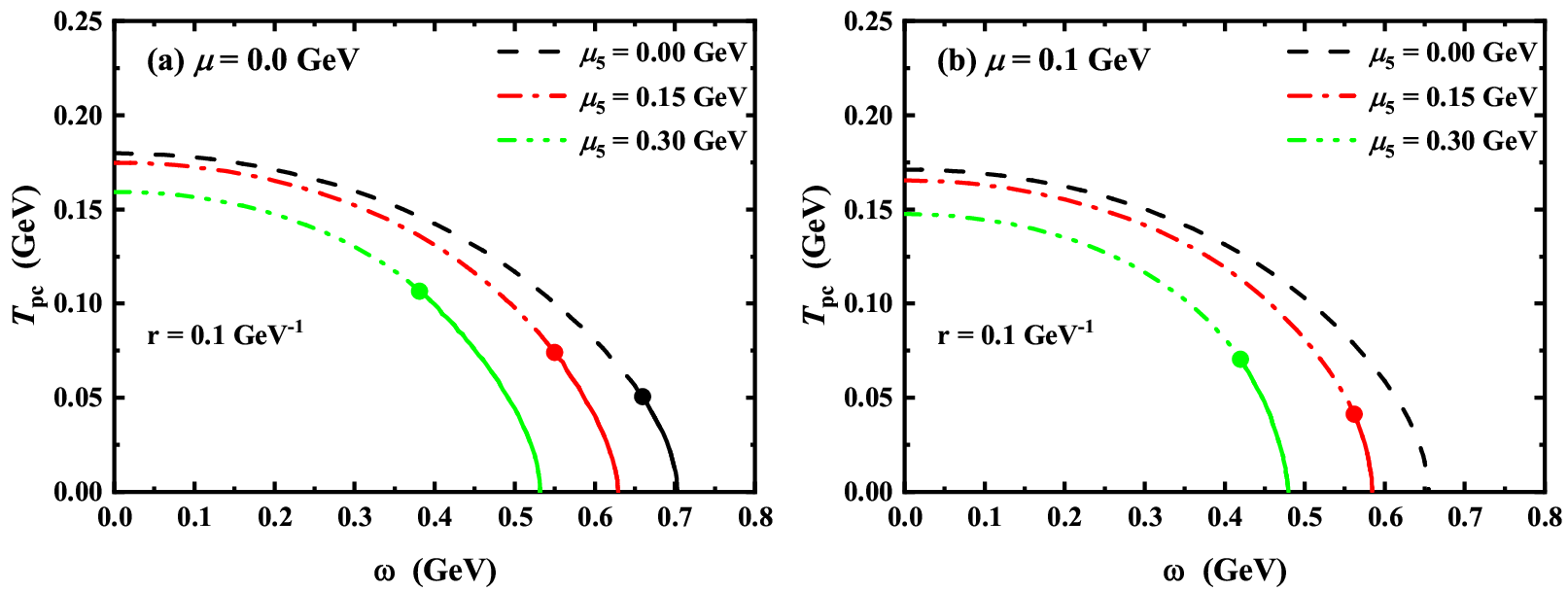}
	\caption{The pseudocritical temperatures ($T_{pc}$) for chiral phase transition as a function of the angular velocity for different chiral chemical potentials,$\mu_{5} = 0$ GeV, $\mu_{5} = 0.15$ GeV and $\mu_{5}=0.3$ GeV at different chemical potentials $\mu = 0$ GeV and $ \mu = 0.1$ GeV, respectively. The solid lines correspond to chiral first-order transition, the full dots correspond to CEP, and the other lines correspond to chiral crossover phase transition.}
	\label{fig:4}
\end{figure}

The critical temperature of the chiral phase transition is determined by the peak of the derivative of the chiral condensate with respect to temperature. As shown in Fig.~\ref{fig:4}, we present the $T_{pc}-\omega$ phase diagrams under different chiral chemical potentials, along with the chemical potential $\mu = 0$ GeV [Fig.~\ref{fig:4}(a)] and $\mu=0.1$ GeV [Fig.~\ref{fig:4}(b)], respectively, where $T_{pc}$ is the pseudocritical temperature. From Fig.~\ref{fig:4}(a) and ~\ref{fig:4}(b), one can see that the rotation effect exhibits a suppressive effect on the phase transition temperature $T_{pc}$. The CEP in a phase transition refers to the point where the first-order phase boundary and the second-order phase boundary intersect on the QCD phase diagram. This point marks a transition in the nature of the phase transition, that is, from a first-order phase transition (with discontinuous physical quantities) to a second-order phase transition (where physical quantities change continuously but have a critical point). It is found that the order of the phase transition in the temperature-angular momentum ($T_{pc} - \omega$) plane changes from a second-order to a first-order phase transition as the angular velocity increases, and the CEP in the $T_{pc} - \omega$ plane is manifested in Fig.~\ref{fig:4}.
We also discussed the impact of  chiral chemical potential  $\mu_{5}$ on the $T_{pc}-\omega$  phase diagram and found that, as the chiral chemical potential $\mu_{5}$ increases, the phase transition temperature $T_{pc}$ decreases, but the critical temperature of the CEP increases.

We also study the characteristics of the $T_{pc} - \omega$ phase diagrams at different chemical potentials. By comparing Figs.~\ref{fig:4}(a) and ~\ref{fig:4}(b), it can be seen that, as the chemical potential $\mu$ increases, the critical temperature of the CEP decreases, while the critical angular velocity increases.
In particular, when $\mu = 0.1$ GeV and $\mu_{5} = 0$ GeV, only a second-order phase transition of the $T_{pc}-\omega$ phase diagrams occurs, with no CEP. In Ref.~\cite{Wang:2018sur}, the chemical potential only lowers the critical temperature of the CEP without altering its critical angular velocity. Here, the effect of the chemical potential $\mu$ on the CEP is opposite to that of the chiral chemical potential $\mu_{5}$.

\subsection{Chirality density}\label{4B}

\begin{figure*}[ht]
	\centering
	\includegraphics[width=\textwidth]{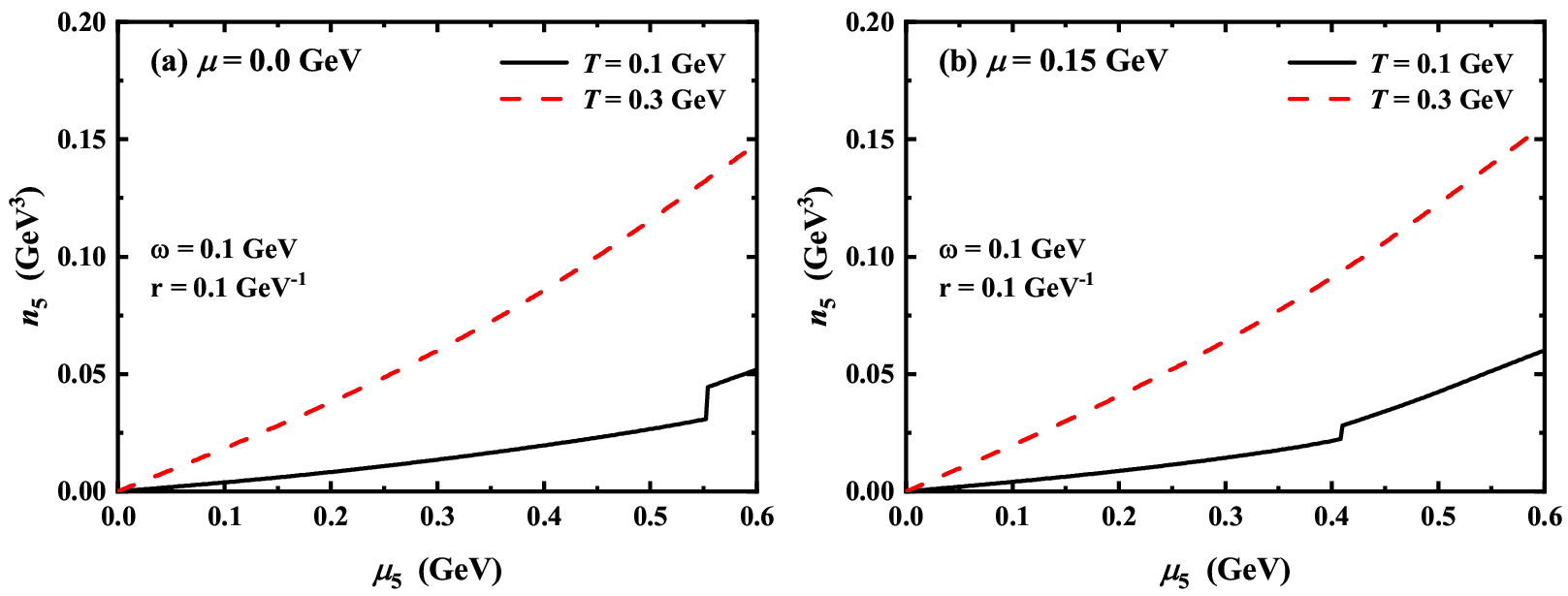}
	\caption{Chirality density $n_{5}$ as a function of chiral chemical potential $\mu_{5}$  for different temperatures $T = 0.1$ GeV and $T = 0.3$ GeV
      at different chemical potentials $\mu = 0$ GeV and $\mu = 0.15$ GeV, respectively.}
	\label{fig:5}
\end{figure*}

The chiral particle number density~\cite{Fukushima:2010bq,Kharzeev:2015znc,Kharzeev:2020jxw,Kharzeev:2007jp} can be denoted as
\begin{equation}\label{eq:20}
	\begin{aligned}
		n_{5}=-\dfrac{\partial\Omega}{\partial\mu_{5}}.
	\end{aligned}
\end{equation}
The dependence of $n_{5}$ on $\mu_{5}$ at different temperatures and different chemical potentials $\mu = 0$ GeV and $\mu=0.15$ GeV is manifested in Figs.~\ref{fig:5}(a) and ~\ref{fig:5}(b), respectively. As shown in Fig.~\ref{fig:5}, when the chiral chemical potential $\mu_{5}$ is zero, the chiral particle number density $n_{5}$ is also zero, and then $n_{5}$ increases with the increase of $\mu_{5}$; that is, the chiral imbalance increases with the increase of chiral chemical potential. In particular, at $T=0.1$ GeV, a jump is observed in the curve, which reflects the discontinuity of the first-order chiral phase transition in relation to $n_{5}(\mu_{5})$. In contrast, at $T=0.3$ GeV, the curve is smooth, indicating that the system is in the chiral symmetric restoration phase. By comparing Figs.~\ref{fig:5}(a) and ~\ref{fig:5}(b), it is found that the relationship between the chiral number density $n_{5}$ and chiral chemical potential $\mu_{5}$ is insensitive to the change in chemical potential $\mu$.

\begin{figure*}[ht]
	\centering
	\includegraphics[width=\textwidth]{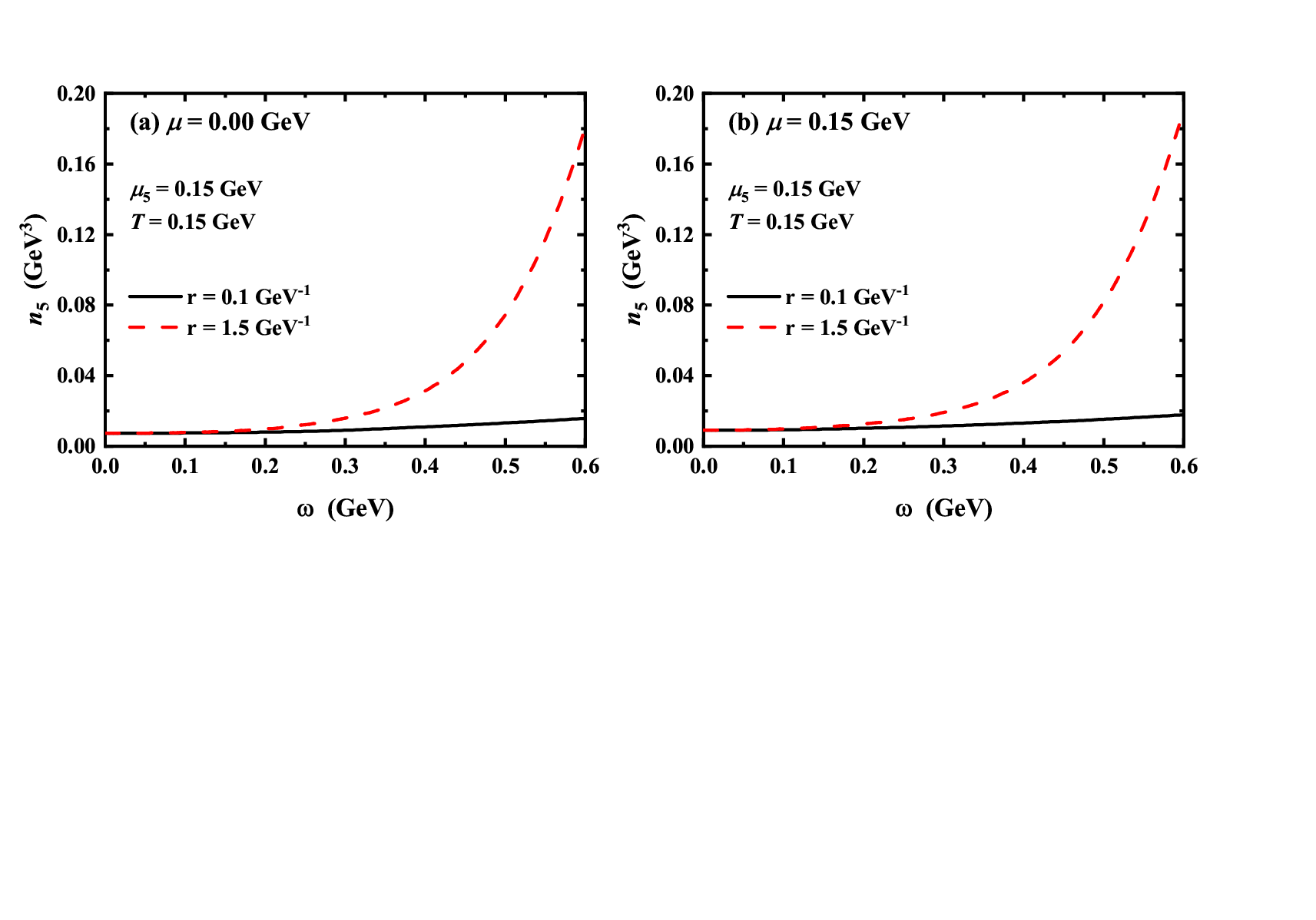}
	\caption{The chirality number density $n_{5}$ as a function of angular velocity $\omega$ for different rotational radii $r = 0.1$ GeV$^{-1}$ and $r = 1.5$ GeV$^{-1}$ with different chemical potentials $\mu = 0$ GeV and $\mu = 0.15$ GeV, respectively.}
	\label{fig:6}
\end{figure*}

Figure~\ref{fig:6} shows that the chiral number density $n_{5}$ increases with the increase of angular velocity $\omega$ at different rotation radii. When the rotation radius is as small as $r = 0.1$ GeV$^{-1}$, this increase is not very obvious, but when the rotation radius reaches as large as $r = 1.5$ GeV$^{-1}$, the chiral number density $n_{5}$ increases significantly with the increase of angular velocity $\omega$. When angular velocity $\omega\geq0.3$ GeV, this increase is particularly significant. It is also found that the change in chemical potential has no significant effect on the dependence of the chirality number density $n_{5}$ on the angular velocity $\omega$.

\subsection{Spin alignment of $\rho$ vector meson}\label{4C}

\begin{figure*}[ht]
	\centering
	\includegraphics[width=\textwidth]{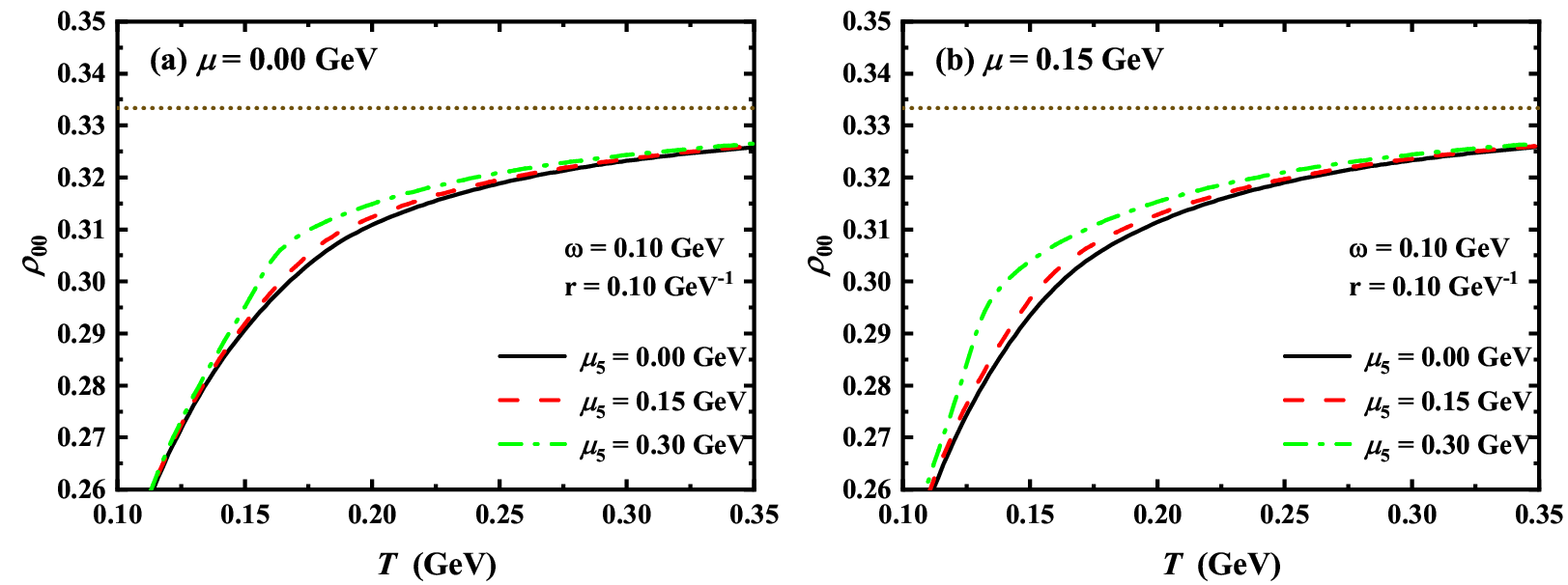}
	\caption{The $\rho$ meson spin alignment as a function of temperature $T$ for different chiral chemical potentials $\mu_{5}=0$ GeV , $\mu_{5}=0.15$ GeV and $\mu_{5}=0.3$ GeV at different chemical potentials $\mu = 0$ GeV and $\mu = 0.15$ GeV, respectively. The gray dotted line is $\rho_{00}=1/3$.}
	\label{fig:7}
\end{figure*}

The temperature-dependent relationship of $\rho_{00}$ under different chiral chemical potentials $\mu = 0$ GeV and $\mu=0.15$ GeV is manifested in Figs.~\ref{fig:7}(a) and ~\ref{fig:7}(b), respectively. The deviation of spin alignment with respect to $1/3$ indicates the polarization of vector mesons along specific directions. It is found that as the temperature rises, $\rho_{00}$ rapidly increases and approaches $1/3$. When $\rho_{00}$ is close to $1/3$, it indicates that the spin alignment of the vector meson $\rho$ is isotropic; that is, the probability is evenly distributed in all spin directions. This means that high temperature makes the spin alignment of the vector meson $\rho$ tend to be isotropic, and it is unlikely that spin polarization happens in a specific direction at high temperature.

Next, we will discuss the influence of chiral chemical potential $\mu_{5}$ on the $\rho_{00}-T$  relationship. An interesting finding is that the chiral chemical potential mainly affects $\rho_{00}$ around the phase transition temperature. It is found that $\rho_{00}$ increases with the chiral chemical potential around the phase transition temperature, and this characteristic is more obvious under a larger chemical potential. At high temperatures ($T\geq0.25$ GeV), the temperature effect dominates, and the influence of the chiral chemical potential on $\rho_{00}-T$ weakens.

\begin{figure*}[ht]
	\centering
	\includegraphics[width=\textwidth]{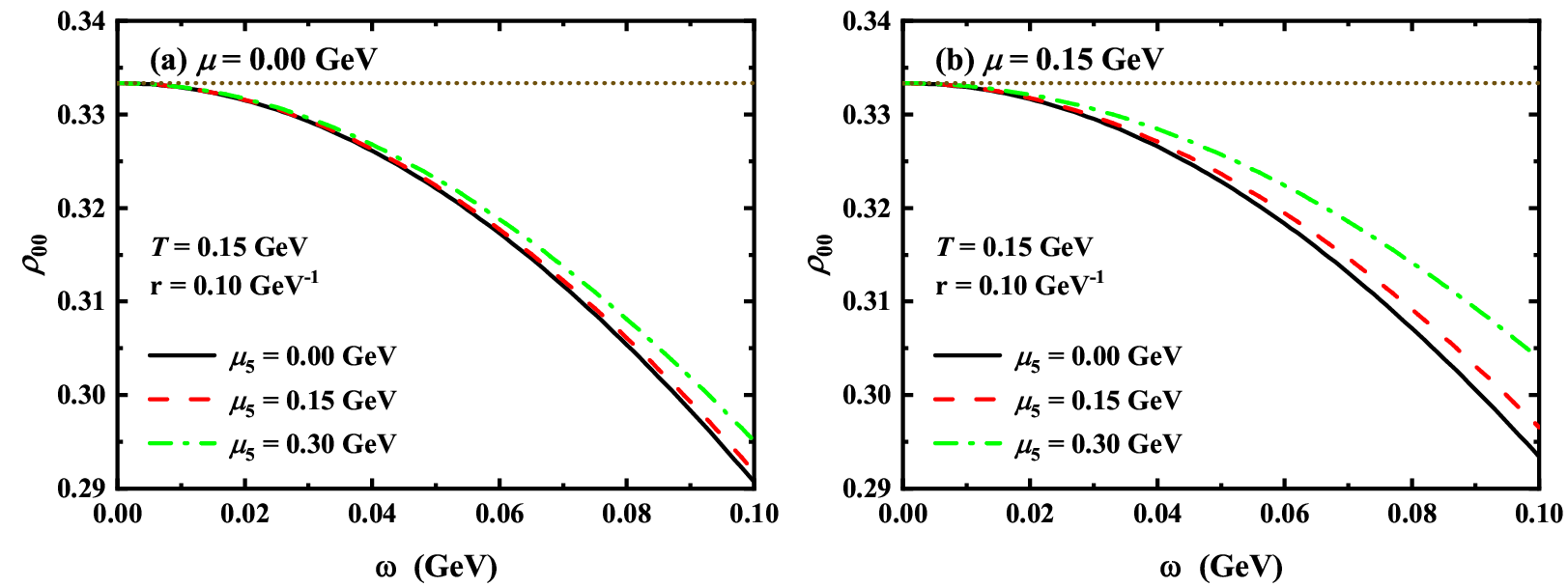}
	\caption{The $\rho$ meson spin alignment as a function of angular velocity $\omega$ for different chiral chemical potentials $\mu_{5}=0$ GeV, $\mu_{5}=0.15$ GeV, and $\mu_{5}=0.3$ GeV at different chemical potentials $\mu = 0$ GeV and $\mu = 0.15$ GeV, respectively. The gray dotted line is $\rho_{00} = 1/3$.}
	\label{fig:8}
\end{figure*}

Reference~\cite{Wei:2023pdf} indicates that $\rho$ meson dissociates at $T = 0.15$ GeV. The angular velocity-dependent relationship of $\rho_{00}$ under different chiral chemical potentials at $T = 0.15$ GeV for $\mu = 0$ GeV and $\mu = 0.15$ GeV is manifested in Figs.~\ref{fig:8}(a) and ~\ref{fig:8}(b), respectively. As can be seen from Fig.~\ref{fig:8}, when $\omega$ is zero, $\rho_{00}$ is close to $1/3$, but as $\omega$ increases, $\rho_{00}$ significantly decreases and deviates $1/3$, indicating that rotation can significantly cause polarization characteristics.

Next, we will discuss the influence of chiral chemical potential on the $\rho_{00}-\omega$ relationship. An interesting finding is that the chiral chemical potential mainly affects $\rho_{00}$ at larger angular velocities. It is found that $\rho_{00}$ increases with the chiral chemical potential at larger angular velocities, and this characteristic is more obvious under a larger chemical potential.

\begin{figure*}[ht]
	\centering
	\includegraphics[width=\textwidth]{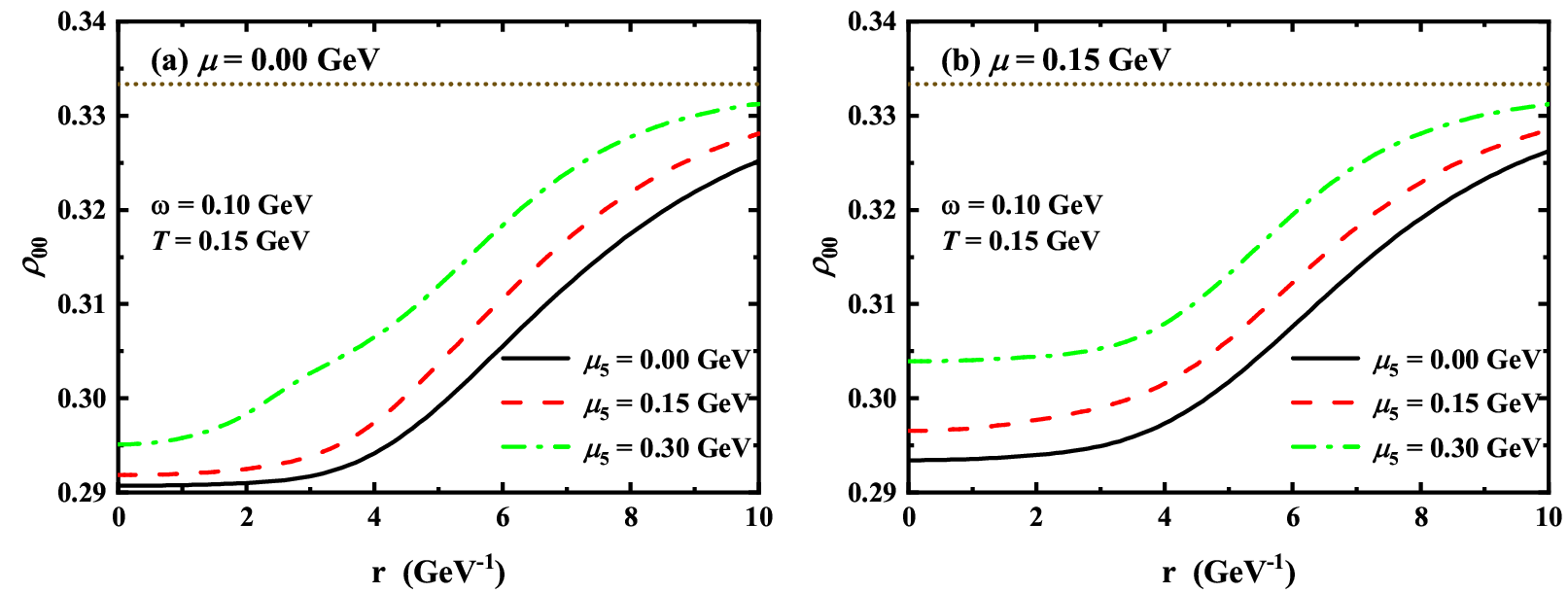}
	\caption{The $\rho$ meson spin alignment as a function of radius $r$ for different chiral chemical potentials $\mu_{5}=0$ GeV, $\mu_{5}=0.15$ GeV, and $\mu_{5}=0.3$ GeV at different chemical potentials $\mu = 0$ GeV and $\mu=0.15$ GeV, respectively. The gray dotted line is $\rho_{00}=1/3.$}
	\label{fig:9}
\end{figure*}

The radius-dependent relationship of $\rho_{00}$ under different chiral chemical potentials for $\mu = 0$ GeV and $\mu = 0.15$ GeV is manifested in Figs.~\ref{fig:9}(a) and ~\ref{fig:9}(b), respectively. As can be seen from Fig.~\ref{fig:9}, $\rho_{00}$ slowly increases with the rotation radius $r$ at smaller radii. However, as the radius increases, $\rho_{00}$ increases rapidly and approaches $1/3$. This indicates that the farther away from the center of rotation, the lower the spin polarization degree of the system. It should be noted that a local approximation method~\cite{Jiang:2016wvv} is utilized to study the radius dependence of spin alignment in the article. Since no boundary effects are introduced, the radius $r$ cannot be interpreted as the system size~\cite{Ebihara:2016fwa}.

Figure~\ref{fig:9} also shows the influence of chiral chemical potential (also known as chiral imbalance) on the rational radius dependence of $\rho_{00}$. Generally speaking, at $T = 0.15$ GeV (near the phase transition temperature), increasing the chiral chemical potential will increase $\rho_{00}$, both near the center of rotation ($r$ is small) and away from the center of rotation ($r$ is large).

\section{CONCLUSIONS AND REMARKS}\label{sec:5}

We work on the two-flavor NJL model under rotation and chiral chemical potential $\mu_{5}$. The influence of chiral imbalance on the $T_{pc}-\omega$ planar chiral phase transition is first studied. Then, the influence of different factors, especially the transverse radius $r$ of the rotating system, on the chiral charge density is also systematically studied. Research shows that as $\mu_{5}$ increases, the critical point of phase transition  will move closer to the $T$ axis. This means that the $T_{pc}-\omega$ phase diagram and phase transition behavior will change under different values of $\mu_{5}$.

In noncentral relativistic heavy-ion collisions, the participating systems have a huge orbital angular momentum, which leads to the production of a strongly vortex field in the QGP. Through spin-orbit interaction, quarks and antiquarks will undergo spin polarization. In the study of the spin alignment of the vector meson $\rho$, $\rho_{00}$ is the $00$ element of the spin density matrix of vector mesons. This value is closely related to the spin alignment characteristics of vector mesons. In summary, the value of $\rho_{00}$ provides important information about the spin state distribution of vector mesons. Approaching $1/3$ of $\rho_{00}$ indicates that the spin alignment is isotropic, while deviating from 1/3 of $\rho_{00}$ indicates the presence of spin polarization, which can help us understand the behavior of strongly interacting matter under extreme conditions.

At high temperatures, $\rho_{00}$ is close to $1/3$, which indicates that the spin alignment of the vector meson $\rho$ is isotropic; that is, the probability is evenly distributed in all spin directions. At lower temperatures, $\rho_{00}$ deviate from $1/3$, which indicates that the spin alignment of the vector meson $\rho$ is nonisotropic and there is a spin polarization phenomenon. It is found that around the phase transition temperature, increasing the chiral chemical potential $\mu_{5}$ can significantly enhances $\rho_{00}$ with finite rotation.

This article employs the quark recombination model to investigate the spin alignment of the vector meson $\rho$. The study reveals that as $\omega$ increases, $\rho_{00}$ significantly decreases and deviates obviously from $1/3$, indicating that rotation can significantly cause polarization characteristics. It is found that at larger angular velocity and larger chemical potential, $\mu_{5}$ can significantly enhance $\rho_{00}$, and make $\rho_{00}$ close to $1/3$.

The study of the $\rho_{00}-r$  relationship near the phase transition temperature is interesting. In general, at $T = 0.15$ GeV (close to the phase transition temperature), $\rho_{00}$ increases with the increase of the chiral chemical potential with different rotational radius. It is found that the farther
away from the center of rotation, the lower the degree of spin polarization of the system.

\begin{acknowledgements}
This work was supported by the National Natural Science Foundation of China (Grants No. 11875178, No. 11475068, and No. 11747115).
\end{acknowledgements}

\bibliography{feng}
\end{document}